\documentclass[aps,prd,superscriptaddress,nofootinbib,amsmath,amsfonts,preprintnumbers,groupedaddress,showpacs,10pt,english]{revtex4-1}
\usepackage{amsmath}
\usepackage{amssymb}
\usepackage{babel}
\usepackage{wrapfig}
\usepackage{cancel}

\usepackage{relsize,exscale}
\makeatletter

\usepackage{array,multirow,graphicx}
\usepackage{dcolumn}
\usepackage{newlfont}
\usepackage{bm}
\usepackage[colorlinks,citecolor=blue,urlcolor=blue,linkcolor=blue]{hyperref}
\usepackage[figtopcap]{subfigure}
\usepackage{color}

\newcommand\be{\begin{equation}}
\newcommand\ba{\begin{eqnarray}}
\newcommand\ee{\end{equation}}
\newcommand\ea{\end{eqnarray}}

\newcommand{\pont}{{\,^\ast\!}R\,R}

\newcommand{\rh}{r_{\text{h}}}

\begin{document}
\title{Spinning (A)dS Black Holes with Slow-Rotation Approximation in Dynamical Chern-Simons Modified Gravity }
\author{  G.  G. L. Nashed$^{1}$ and S. Capozziello$^{2,3,4}$ }

\affiliation {$^1$Centre for Theoretical Physics, The British University in Egypt, P.O. Box 43,\\ El Sherouk City, Cairo 11837, Egypt\\
$^2$Dipartimento di Fisica ``E. Pancini``, Universit\'a di Napoli ``Federico II'',
Complesso Universitario di Monte Sant' Angelo, Edificio G, Via Cinthia, I-80126, Napoli, Italy\\
$^3$ Istituto Nazionale di Fisica Nucleare (INFN),  Sezione di Napoli,
Complesso Universitario di Monte Sant'Angelo, Edificio G, Via Cinthia, I-80126, Napoli, Italy\\
$^4$Scuola Superiore Meridionale, Largo S. Marcellino 10, I-80138, Napoli, Italy.}

\date{\today}

\begin{abstract}

One of the most crucial areas of gravity research, after the direct observation of gravitational waves, is the possible modification of General Relativity at ultraviolet and infrared scales. In particular, the possibility of parity violation should be considered in strong field regime.  The Chern-Simons  gravity takes into account parity violation in strong gravity regime.  For all conformally flat spacetimes and spacetimes with a maximally symmetric two-dimensional subspace, Chern-Simons gravity is identical to General Relativity. Specifically, the Anti-de Sitter (A)dS-Kerr/Kerr black hole  is not a solution for Chern-Simons gravity. { The slow-rotating BH and
the quadratic order in spin solutions are some of the known solutions to quadratic order in spin and they  are rotating solutions in the frame of dynamical Chern-Simons gravity}.
 In the present study, for the (A)dS slow-rotating situation (correct to the first order in spin), we derive the linear perturbation equations controlling the metric and the dynamical Chern-Simons field equation corrected to the linear order in spin and to the second order in the Chern-Simons coupling parameter. We show that the black hole of the (A)dS-Kerr solution is stronger  (i.e. more compact and energetic) than the  Kerr black hole solution and the reason for this feature comes form  contributions at Planck scales. Moreover, we calculate the thermodynamical quantities related to this black hole. Finally, we calculate the geodesic equation  and derive the effective potential of the black hole. We show that as the numerical value of the rotation parameter increases, there is a peak, and as the rotation parameter increases further, the peak becomes positive, preventing  the photons from outside  to fall into the black hole.

\end{abstract}





\maketitle

\section{Introduction}
\label{intro}

Experiments and observations over the last few decades  revealed several phenomena that cannot be explained by General Relativity (GR) at  ultraviolet and infrared scales. Most of these issues, demonstrate GR incompatibilities at cosmological and astrophysical scales. On the one hand, GR fixes observations at solar system scales perfectly. It is, however, incapable of providing an exhaustive and self-consistent picture of phenomena such as late-time universe acceleration, which has previously been attributed to the nebulous concept of dark energy. Similarly, inconsistencies in galaxy rotation curves are attributed to dark matter, which has never been detected under the standard of a new fundamental particle out of the Standard Model. These are just two examples, but the issues that arise when the theory is applied to large-scale structure observations are numerous~\cite{Joyce:2014kja,Koyama:2015vza,Faraoni:2010pgm}. In the frame of these concerns, modified theories of GR have been constructed to describe the gravitational interaction by incorporating extra-terms or alteratives into the Einstein-Hilbert action.  The $f(R)$  gravity, for example, is a straightforward  extension where  the starting action is a general function of the Ricci scalar curvature~\cite{Capozziello:2002rd,Capozziello:2009nq, Nojiri:2017qvx,Capozziello:2011et,Capozziello:2015hra,Ribeiro:2021gds,Capozziello:2021wwv}. Usually, functions of higher-order curvature invariants can serve the role of the Ricci scalar in the action, assuming extensive interpolation at the small-scale regime, i.e.,  ultraviolet scales~\cite{Capozziello:2014ioa, Terrucha:2019jpm, Barros:2019pvc, Bajardi:2020mdp, Blazquez-Salcedo:2016enn, Blazquez-Salcedo:2017txk}. Several outstanding issues, concerning the early and late-time acceleration of the universe, can be addressed by coupling  geometry to a scalar field $\phi$, as  discussed, for example,  in Refs.~\cite{Amendola:1999qq, Uzan:1999ch, Halliwell:1986ja, Bajardi:2020xfj}. Modified actions produce effective energy-momentum tensors of the gravitational field,  similar to the phenomenology adopted for dark matter and dark energy~\cite{Capozziello:2007ec, Clifton:2011jh, Bamba:2012cp, Nojiri:2017ncd, Capozziello:2019klx,Capozziello:2018ddp}.

GR cannot be considered at small scales using the same criteria as the other field theories. Indeed, this theory results in several flaws in view of  a self-consistent  theory of quantum gravity. Specifically,  to equip   GR with the formalism of other Standard Model interactions, the former must be recast as a Yang-Mills theory. Furthermore, even the 2-loop expansion of  gravitational action demonstrates that incurable divergences occur in GR: This means that  it cannot be renormalized using the standard regularization procedure. Despite numerous attempts to integrate GR and Quantum Field Theory, a complete theory of quantum gravity remains elusive at the moment.  For example, the ``Arnowitt-Deser-Misner"  (ADM) formalism, which deals with an infinite dimensional superspace, leads to a Schr\"odinger-like equation known as the Wheeler-De-Witt equation~\cite{Bajardi:2020fxh, Hartle:1983ai, Hawking:1983hj}. The ADM formulation fails to account for a complete quantum theory of gravity and has several flaws that cannot be overcome also if its use in quantum cosmology gives interesting results.

Because gauge theories are currently the only candidates capable of selecting renormalizable quantum field theories, a forthcoming theory that aims to address high-energy issues must cope with gravitational interaction as a gauge theory. The so-called teleparallel gravity is an example of  gauge theory of gravity dealing with a flat tangent space-time, which is invariant under the local translation group and whose action differs from Einstein-Hilbert by  a total divergence. It describes gravity as a torsional spacetime, including the antisymmetric contribution of connections. The gravitational field is represented by vielbiens (tetrads in four dimensions), and the affinities associated with them are represented by the Weitzenb\"ock connection. For a  discussion on teleparallel gravity and its applications, see e.g. Refs.~\cite{Cai:2015emx,Ferraro:2006jd,Arcos:2004tzt, Bajardi:2021tul, Capozziello:2022zzh}.

In 1971, Lovelock proposed an additional generalization of GR~\cite{Lovelock:1971yv}. The Lovelock-Zumino Lagrangian (or simply the Lovelock Lagrangian) is the most general torsionless Lagrangian in such a theory, leading to second-order field equations. In four dimensions, the Gauss-Bonnet term appears, but it does not contribute to the equations of motion. In fact, in four dimensions, the Gauss-Bonnet term becomes a topological surface term, while, in five-dimensions, it becomes non-trivial. By construction, any Lovelock Lagrangian, regardless of dimension, is always invariant, at least under the local Lorentz group. Certain combinations of the coupling constants, however, make the theory invariant for other gauge groups. The Lovelock three-dimensions Lagrangian is an exception, as it is invariant under the local Poincare group for any coupling parameter. Furthermore, it proves that there is a subclass of Lovelock Lagrangians where the coupling constants are coupled in some way such that the Lagrangian external derivative offers a topological invariant. Such Lagrangians are known as "Chern-Simons Lagrangians," and they contribute significantly to dynamics only in odd dimensions.

Chern-Simons (CS) improved gravity is a well-known four-dimensional scalar-tensor theory which was first proposed in Ref.~\cite{Jackiw:2003pm}. It is inspired by anomaly cancellation in curved spacetimes, particle physics, and string theory low-energy limit (for a review, see~\cite{Alexander:2009tp}). The theory has a non-minimal coupling between a scalar field and the Chern-Pontryagin density. This interaction, in particular, encapsulates parity-violating characteristics in the strong gravity regime. Furthermore, its equations of motion effectively reduce to those of topologically massive gravity under certain conditions~\cite{Deser:1981wh,Deser:1982vy}, that is a three-dimensional gravity built from CS theory. Because the CS coupling generates higher-order field equations, this theory does not belong to the Horndeski family of theories~\cite{Horndeski:1974wa}. In order to avoid the drawbacks associated with the Cauchy initial-value problem, it should be regarded as an effective theory derived from the ultra-violet completion of GR~\cite{Delsate:2014hba}.

In the Einstein-Hilbert action plus a new parity-violating term, four-dimensional correction identifies the action for CS-modified gravity. When it was discovered that string theory requires just such a correction to remain mathematically consistent, interest in the model skyrocketed. The Green-Schwarz anomaly canceling mechanism in the perturbative string sector requires such a correction upon four-dimensional compactification. In general, due to duality symmetries, such a correction arises in the presence of Ramond-Ramond scalars.

In particular, a slowing rotating black hole (BH) can be adorned with (secondary) scalar hair in this parity-violating axion field.  An axion field hair that dresses a slowly rotating BH, as the result of  axion field coupling to a Lorentz CS term, is reported in the papers  \cite{Campbell:1990ai,Campbell:1991kz}. This finding suggests that non-minimal gravitational couplings could lead to novel effects in BH backgrounds. This solution was extended in \cite{Duncan:1992vz}, where it was discovered that the charges presented by the axion hair are defined by the background rotating BH mass, angular momentum, and gauge charges. In the small coupling slow rotation limit, a solution describing a rotating BH was found by allowing the axion field to be dynamical \cite{Yunes:2009hc}. In dynamical CS modified gravity, static and rotating black string solutions were investigated \cite{Cisterna:2018jsx,Corral:2021tww}. { In the framework of dynamical Chern-Simons and Einstein-dilaton-Gauss-Bonnet theory \cite{Maselli:2017kic}, BH solutions that differ from the Kerr solution are presented. Using the so-called extremal limit, rotating black hole solutions based on dynamical Chern-Simons gravity were derived in \cite{McNees:2015srl}.} It is the aim of the present research to generalized the previous studies and try to derive an (A)dS-Kerr BH solution in the framework of dynamical CS gravity.

The outline of the present  paper is the following.
In Sec.~\ref{ABC}, we review the basics of CS modified gravity.
 Sec.~\ref{axisym} is devoted to   dynamics  of a (A)dS-slowly rotating BH  with small
CS coupling constants. In Sec.~\ref{properties},  we  study  physical properties of the
 (A)dS-Kerr BH solution by calculating its thermodynamical quantities in view of possible   astrophysical applications.   In Sec.~\ref{geo}, we calculate the geodesic equation of this BH and derive its effective potential showing that the increasing of  rotation parameter makes the peak  positive, and thus  prevents the photon  coming from outside to fall into the BH. In
Sec.~\ref{conclusions},  we summarize the results   and discuss  possible future researches.

The following notation is adopted: In four-dimensional  spacetime,  the  signature is 
$(-,+,+,+)$~\cite{Misner:1973prb}, where Latin symbols $(a,b,\ldots,h)$
refer to spacetime indices. Square and round brackets   denote 
anti-symmetrization and symmetrization respectively, i.e.,  $T_{[ab]}=\frac12 (T_{ab}-T_{ba})$ and  $T_{(ab)}=\frac12
(T_{ab}+T_{ba})$.  Commas are used for  partial derivatives   (e.g.~$\partial
\varphi/\partial r=\partial_r\varphi=\varphi_{,r}$).  
 We adopt the Einstein summation  and geometrized units with $G=c=1$.

\section{The Chern-Simons modified gravity}
\label{ABC}
Let us give now some  basic notions  of CS modified gravity~\cite{Alexander:2009tp}. It is well known that CS gravity can be divided into  two main classes:  Non-dynamical and dynamical models. Non-dynamical cases are not interesting because they  give no new physics. In particular, they do not give solutions different from the Schwarzschild one, therefore we will not discuss them anymore.  On the other hand, we  will discuss  dynamical cases below.\\

\centerline{\bf Basics}

Let us start with the following action:
\be
\label{CSaction}
S = S_{ EH} + S_{ CS} +  S_{\varphi} + S_{ mat},
\ee
where
\ba
\label{EH-action}
S_{{EH}} &=& \kappa \int_{{\cal{V}}} d^4x  \sqrt{-g}  \left(R-2\Lambda\right),
\qquad \qquad \qquad \qquad
S_{{CS}} = \frac{\alpha}{4} \int_{{\cal{V}}} d^4x  \sqrt{-g} \;
\varphi \; \pont\,, \nonumber
\\
\label{Theta-action}
S_{\varphi} &=& - \frac{\beta}{2} \int_{{\cal{V}}} d^{4}x \sqrt{-g} \left[ g^{\alpha \beta}
\left(\nabla_{\alpha} \varphi\right) \left(\nabla_{\beta} \varphi\right) + 2 V(\varphi) \right], \qquad \qquad
S_{\textrm{mat}}= \int_{{\cal{V}}} d^{4}x \sqrt{-g} {\cal{L}}_{\textrm{mat}}.\nonumber
\ea
The first term in Eq.~\eqref{CSaction} is the  Einstein-Hilbert  with $\Lambda$ is the cosmological constant that can be written as ${\displaystyle \Lambda=-\frac{3}{l^2}}$ where $l$ is the Planck scale.  The second one is the CS term while the third contribution is due to the scalar-field.
The last term is an additional matter sources where ${\cal{L}}_{\textrm{mat}}$
is the matter Lagrangian density. Here we use the
following notation: $\kappa^{-1} = 16 \pi G$. The symbols $\alpha$ and $\beta$ are {{dimensional}} coupling constants, $\nabla_{\alpha}$ is the covariant derivative w.r.t. the metric tensor $g_{\alpha \beta}$,
 $g$ is the determinant of the metric,  and $R$ is the Ricci scalar. The expression $\pont$ is the Pontryagin density, figured as:
\be
\label{pontryagindef}
\pont= R \tilde R = {\,^\ast\!}R^\alpha{}_\beta{}^{\gamma \delta} R^\beta{}_{\alpha \gamma \delta}\,.
\ee
The  dual Riemann tensor is:
\be
\label{Rdual}
{^\ast}R^\alpha{}_\beta{}^{\gamma \delta}=\frac12 \epsilon^{\gamma \delta \rho \rho_1}R^\alpha{}_{\beta \rho \rho_1}\,,
\ee
where $\epsilon^{\gamma \delta \rho \rho_1}$ is the 4-dimensional Levi-Civita tensor.

The spacetime function $\varphi$  is  the {{CS coupling field}}. It parameterizes deformations from GR.  The Pontryagin density is equal to the total divergence of the CS topological current $K^{a}$  therefore, if $\varphi= \textrm{constant}$, the CS modified gravity reduces to GR. It is
\be
\nabla_\alpha K^\alpha = \frac{1}{2} \pont,
\label{eq:curr1}
\ee
where
\be
K^\alpha =\epsilon^{\alpha \beta \gamma \delta} \Gamma^\rho_{\beta\rho_1} \left(\partial_\gamma\Gamma^{\rho_1}_{\delta \rho}+\frac{2}{3} \Gamma^{\rho_1}_{\gamma \rho_2}\Gamma^{\rho_2}_{\delta \rho}\right)\,,
\label{eq:curr2}
\ee
and $\Gamma$ is the Christoffel connection. Eq. \eqref{eq:curr2} can be used  to rewrite $S_{\textrm{CS}}$ in  the form:
\be
\label{CS-action-K}
S_{\textrm{CS}} = \alpha
\left( \varphi \; K^{\alpha} \right)|_{\partial {\cal{V}}}
-
 \frac{\alpha}{2} \int_{{\cal{V}}} d^4x  \sqrt{-g} \;
\left(\nabla_{\alpha} \varphi \right) K^{\alpha}.
\ee
The first term, corresponding to the CS correction, is typically ignored since it is calculated on the  spacetime boundary~\cite{Grumiller:2008ie}.
The variation of  action \eqref{CSaction} w.r.t.  the metric and  the CS  field gives the field   equations:
\ba
\label{eom}
&&R_{\alpha \beta} -2g_{\alpha \beta} \Lambda+ \frac{\alpha}{\kappa} C_{\alpha \beta} = \frac{1}{2 \kappa} \left(T_{\alpha \beta} - \frac{1}{2} g_{\alpha \beta} [T -4\Lambda] \right),\qquad \qquad  \beta\; \square \varphi = \beta \; \frac{dV}{d\varphi} - \frac{\alpha}{4} \pont,
\ea
with $R_{\alpha \beta}$  being the Ricci second order tensor and $\square = \nabla_{a} \nabla^{a}$ is the d'Alembert operator.
Here the expression $C_{\alpha \beta}$ is  the C-tensor defined as:
\be
\label{Ctensor}
C^{\alpha \beta} = w_\rho
\epsilon^{\rho \gamma \delta(\alpha}\nabla_\delta R^{\beta)}{}_\gamma+w_{\gamma \delta}{\,^\ast\!}R^{\gamma(\alpha \beta)\delta}\,,
\ee
where
\be
\label{v}
w_\alpha=\nabla_\alpha\varphi\,,\qquad
w_{\alpha \beta}=\nabla_\alpha\nabla_\beta\varphi.
\ee
The total stress-energy tensor is:
\be\label{Tab-total}
T_{\alpha \beta}= T^{\textrm{mat}}_{\alpha \beta} + T_{\alpha \beta}^{\varphi},
\ee
where $T^{\textrm{mat}}_{\alpha \beta}$ is given by
standard matter sources,
and $T_{\alpha \beta}^{\varphi}$ is the scalar field contribution defined as:
\be
\label{Tab-theta}
T_{\alpha \beta}^{\varphi}
=   \beta  \left[  \left(\nabla_{\alpha} \varphi\right) \left(\nabla_{\beta} \varphi\right)
    - \frac{1}{2}  g_{\alpha \beta}\left(\nabla_{\alpha} \varphi\right) \left(\nabla^{\alpha} \varphi\right)
-  g_{\alpha \beta}  V(\varphi)  \right].
\ee
If the equation of motion of  scalar field  $\varphi$ holds then the strong Equivalence Principle ($\nabla_{\alpha} T^{\alpha\beta}_{\textrm{mat}} = 0$) is verified in CS modified gravity. This is due to the fact that when one considers the divergence of Eq.~\eqref{eom}, the first and second terms  on the LHS vanishes by the Bianchi identities. However, the third term is proportional to the Pontryagin density by:
\be
\label{nablaC}
\nabla_\alpha C^{\alpha \beta} = - \frac{1}{8} w^\beta \pont.
\ee

\section{(A)dS-Rotating Black Holes in Dynamical Chern-Simons Modified Gravity }\label{axisym}

In this section, we are going to study (A)dS-rotating BHs in  dynamical  Chern-Simons  modified gravity.
Without using any approximation, it is difficult to analyze the stationary and axi-symmetric line-element in dynamical CS gravity.   Therefore,  we adopt a couple of approximations  and solve the field equations up to second order in  the
perturbative expansion.

\subsection{The approximation structure}
\label{approx}
{ We use two approximations structures: a {{\it small-coupling}}  and a {{\it slow-rotation}} approximations, for more details of this analysis \citep[see][]{Yunes:2009hc}.}
In the first structure,  we use the CS modification as a small distortion of GR. It permits us to adopt the following metric decomposition (up to second order):
\be
g_{\alpha \beta} = g_{\alpha \beta}^{(0)} + \zeta g^{(1)}_{\alpha \beta}(\varphi) + \zeta^{2} g^{(2)}_{\alpha \beta}(\varphi).
\label{small-cou-exp0}
\ee
Here $g_{\alpha \beta}^{(0)}$ is the background metric  which satisfies the Einstein field equations, like Kerr metric, whilst $g_{\alpha \beta}^{(1)}(\vartheta)$ and $g_{\alpha \beta}^{(2)}(\vartheta)$ are the first and the second-order CS perturbations which depend on $\varphi$. The parameter $\zeta$ refers to the order of  small-coupling approximation.

On the other hand, by  slow-rotation procedure, we are able to re-expand the background and the $\zeta$-perturbations in powers of the Kerr rotation parameter $a$.  Consequently, the background metric and the metric perturbation become:
\ba
\label{small-cou-exp}
g_{\alpha \beta}^{(0)} &=& \eta_{\alpha \beta}^{(0,0)} + \epsilon \; h_{\alpha \beta}^{(1,0)} + \epsilon^{2} h_{\alpha \beta}^{(2,0)},
\nonumber \\
\zeta g_{\alpha \beta}^{(1)} &=& \zeta h_{\alpha \beta}^{(0,1)} + \zeta \epsilon \; h_{\alpha \beta}^{(1,1)} + \zeta \epsilon^{2} h_{\alpha \beta}^{(2,1)},
\nonumber \\
\zeta^{2} g_{\alpha \beta}^{(2)} &=& \zeta^{2} h_{\alpha \beta}^{(0,2)} + \zeta^{2} \epsilon \; h_{\alpha \beta}^{(1,2)} + \zeta^{2} \epsilon^{2} h_{\alpha \beta}^{(2,2)},
\ea
where $\epsilon$ refers to the order of the slow-rotation expansion. We emphasize that the notation $h^{(a,b)}_{\alpha \beta}$ refers to terms of ${\cal{O}}(a,b)$, which refers to  ${\cal{O}}(\epsilon^{a})$ and ${\cal{O}}(\zeta^{b})$. As an example, in Eq.~\eqref{small-cou-exp}, $\eta_{\alpha \beta}^{(0,0)}$ is consider as the background metric when $a\rightarrow 0$, whilst $h_{\alpha \beta}^{(1,0)}$ and $h_{\alpha \beta}^{(2,0)}$ refer to the background metric  of the first and second-order expansions  in the spin parameter.

By combining the two approximate methods, we are able to create a bivariate expansion with two independent parameters $\zeta$ and $\epsilon$, which is provided, at second order, as:
\be
g_{\alpha \beta} = \eta_{\alpha \beta}^{(0,0)} + \epsilon h_{\alpha \beta}^{(1,0)} + \zeta h_{\alpha \beta}^{(0,1)} + \epsilon \zeta h_{\alpha \beta}^{(1,1)} + \epsilon^{2} h_{\alpha \beta}^{(2,0)} + \zeta^{2} h_{\alpha \beta}^{(0,2)},
\ee
When we deal with first-order terms, we  mean those that are ${\cal{O}}(1,0)$ or ${\cal{O}}(0,1)$, but when we  deal with second-order terms, we mean those that are ${\cal{O}}(2,0)$, or ${\cal{O}}(0,2)$ or ${\cal{O}}(1,1)$.

The   parameters $\epsilon$ and $\zeta$ have a key role into this discussion.  Because the slow-rotation technique expands the Kerr parameter, thus its {\it{dimensionless}} expansion parameter must be $\alpha/M$. Consequently, a term in the equations multiplied by $\epsilon^n$ is of the form ${\cal{O}}\left((\alpha/M)^n\right)$.
The small-coupling expansion should depend on the ratio of CS coupling to the GR coupling, i.e., $\alpha/\kappa$, since this combination multiplies the C-tensor in Eq.~\eqref{eom}.

The small-coupling strategy creates a clear iteration or boot-strapping scheme by combining with the structure of the modified field equations. One can observe that  the source of the $\varphi$-evolution equation is always of a lower order than the CS correction to the Einstein equations from Eq.~\eqref{eom}. According to this observation, one can independently solve the evolution equation for $\varphi$ first. In order to determine the CS correction to the metric, the solution  of $\varphi$ can then be applied to the modified field equations.
This method can then theoretically be repeated in order to solve higher order  expansion parameters.

Let us provide  an example of a boot-strapping scheme.  To make this easier, let us choose, for the moment,  units  so that $\varphi$   is dimensionless and  $\beta = \kappa$. Then $\alpha$ controls the small-coupling expansion parameter only via $\zeta = {\cal{O}}[\alpha^{2}/(\kappa^{2} M^{4})]$.  The RHS  of Eq.~\eqref{eom}, in these units, is proportional to $\zeta^{1/2}$, but the second term in Eq.~\eqref{eom} is proportional to $\zeta$, which suggests that $\varphi$ is a Frobenius series with a fractional structure, that is
\be
\varphi = \zeta^{1/2} \sum_{n=0}^{\infty} \zeta^{n} \varphi^{(n)},
\ee
while, as required by Eq.~\eqref{small-cou-exp0}, the metric perturbation is a regular series in natural powers of $\zeta$.

Alternatively, various $\varphi$ units  might be used which could  slightly alter the order of counting. As an example, let us  choose units, 
which, by a  dimensional analysis yields $[\alpha] = L^{4}$ and $[\varphi] = L^{-2}$.  It gives $\zeta = {\cal{O}}(\alpha/M^{4})$, but the RHS of Eq.~\eqref{eom} is  proportional to $\zeta^{0}$  up to the leading order.  Using these units make $\vartheta$ and $g_{\alpha \beta}$ to  have expansions in natural powers of $\zeta$, while the leading-order expansion of the former is $1/\zeta$ much larger than the latter.

The $\varphi$-evolution equation  is then seen to always be of lower order in comparison  to the modified field equation,  leading to a clear boot-strapping strategy regardless of the units.  A term of the form ${\cal{O}}(1,1)$ or ${\cal{O}}(1,1/2)$ leads  to $(\alpha/M) (\alpha/\beta)$ depending  on the choice of $\beta$. We shall assume that $\beta \propto \alpha$, for the  sake of order counting, but  we will leave all factors of $\beta$ explicit. By using this choice, the parameter ratio is of order unity and both $\varphi$ and $g_{\alpha \beta}$ have expansions  $\zeta$ in natural powers.

This boot-strapping procedure is similar to the one called semi-relativistic approximation~\cite{Ruffini:1981af}, where one models  extreme-mass ratio inspirals by resolving the geodesic equations and ignoring the self-force of the particle. Even through this approximation, one cannot  solve the field equations, if the background is too complicated (like, the Kerr metric). Therefore, the small-rotation procedure,  introduced above, gives independent equations, obtained from the boots trapping procedure in the small-coupling approximation.  This combined procedure allows  to solve the equations in exact way.

The following remark should be taken into account when we use the bivariate expansions:
Both   $\epsilon$ and  $\zeta$ are {{independently}} small. The only constrain we he have to consider is  that $\epsilon$ is not  proportional to an inverse power of $\zeta$, because this could violate the above requirement. Also, we have to stress that, as it is well known in perturbation theory, $\zeta$  and $\epsilon$   are only parameters and are not equal to $\alpha/M$ or $\xi/M^{4}$, instead they multiply terms in the output
equations of the same  order. Here $\xi$ is a distortion of the (A)dS-Kerr metric which will be defined below.
 Because the parameters $\epsilon$ and  $\zeta$ do not have any physical meaning, thus we can  set them to unity at the end of the calculation at a given order of perturbation, as we will show below.

\subsection{(A)dS-slowly rotating black hole solutions}\label{slow-rot}

The expansion of slow-rotation, using background metric, can be constructed through   the Hartle-Thorne approximation~\cite{Thorne:1984mz,Hartle:1968si}, where the line element is written in  Boyer-Lindquist coordinates,  $(t,r,\theta,\phi)$,  as:
\ba
\label{slow-rot-ds2}
ds^{2} &=& - k  \left[1 + s(r,\theta)\right] dt^{2}
+ \frac{1}{k}  \left[1 + p(r,\theta)\right] dr^{2}
+r^{2} \left[1 + q(r,\theta) \right] d\theta^{2}
\nonumber \\
&+& r^{2} \sin^{2}{\theta} \left[1 + n(r,\theta) \right] \left[ d\phi  - \omega(r,\theta) dt \right]^{2},
\ea
where ${\displaystyle k = \frac{r^2}{l^2}+1 - \frac{2 m}r}$  is the (A)dS-Schwarzschild spacetime with $M$ being the BH mass and $l$ is a length related to the cosmological constant.
Here $s(r,\theta)$, $p(r,\theta)$, $q(r,\theta)$, $n(r,\theta)$ and $\omega(r,\theta)$ are the perturbations.

We have written the metric of Eq.~\eqref{slow-rot-ds2} as given  in~\cite{Thorne:1984mz,Hartle:1968si}, however,  the metric perturbations should be expanded like  a series using $\zeta$ and $\epsilon$. The metric perturbations up to the second order yield:
\ba
 s(r,\theta) &=& \epsilon \; s_{(1,0)} + \epsilon \;  \zeta \; s_{(1,1)} + \epsilon^{2} \; s_{(2,0)},
 \nonumber \\
  p(r,\theta) &=& \epsilon \; p_{(1,0)} + \epsilon \;  \zeta \; p_{(1,1)} + \epsilon^{2} \; p_{(2,0)},
 \nonumber \\
  q(r,\theta) &=& \epsilon \; q_{(1,0)} + \epsilon \;  \zeta \; q_{(1,1)} + \epsilon^{2} \; q_{(2,0)},
 \nonumber \\
  n(r,\theta) &=& \epsilon \; n_{(1,0)} + \epsilon \;  \zeta \; n_{(1,1)} + \epsilon^{2} \; n_{(2,0)}.
 \nonumber \\
  \omega(r,\theta) &=& \epsilon \; \omega_{(1,0)} + \epsilon \;  \zeta \; \omega_{(1,1)} + \epsilon^{2} \; \omega_{(2,0)}.
 \ea
In  Eq.~\eqref{slow-rot-ds2}, we have not taken into account  terms of ${\cal{O}}(0,0)$ since they are  involved  in the (A)dS-Schwarzschild spacetime.
Moreover, we { assume} that, when the Kerr spin parameter is $a\rightarrow 0$, we recover (A)dS-Schwarzschild as a trivial solution,
involving that all terms of order ${\cal{O}}(0,n)$ are vanishing.
 Therefore, the CS correction term should be  linear in the Kerr spin parameter $a$.
We can read-off the metric perturbations, proportional to $\zeta^{0}$, from the slow-rotation limit of the Kerr metric in GR  up to the first order :
\ba
s_{(1,0)} &=& p_{(1,0)} = q_{(1,0)} = n_{(1,0)} =0,
\nonumber \\
\omega_{(1,0)} &=& \left(\frac{r^{2}}{l^2}-\frac{2m}{r}\right) a
\ea
and to the second order:
\ba
s_{(2,0)} &=& -\frac{\left(\frac{r^{2}}{l^2}\sin^2\theta+\frac{2m}{r}\cos^2\theta\right)a^2}{kr^{2}}\nonumber\\
p_{(2,0)} &=& \frac{a^{2}\left[r(1+\frac{r^2}{l^2})\sin^2\theta+\frac{2m}{r}\cos^2\theta\right]}{\frac{r^{2}}{l^2}k}\,,
\nonumber\\
q_{(2,0)} &=& \frac{a^{2}}{r^{2}} \left(1+\frac{r^2}{l^2}\right)\cos^{2}{\theta},
\nonumber \\
n_{(2,0)} &=& -\frac{a^{2}}{r^{2}} \left(1+\frac{r^2}{l^2} + \frac{2 m}{r} \sin^{2}{\theta} \right),
\qquad \qquad
\omega_{(2,0)} = 0\,.
\ea
All fields are expanded in small-coupling and slow-rotation approximation, inclusive of the CS coupling field. Using the evolution equation, given by the second term of Eq.~\eqref{eom}, we can examine the leading-order behavior
of $\varphi$. From the second term of Eq.~\eqref{eom}, we get that $\partial^{2} \varphi \sim (\alpha/\beta) \pont$, since the Pontryagin density has a zero value up to order $a/m$.
Therefore, the CS scalar field  leading order must be $\varphi \sim (\alpha/\beta) (a/m)$, which  is proportional to $\epsilon$.
Moreover, since the (A)dS-Schwarzschild metric is the only solution up to zero-angular
momentum limit, therefore, we should have  $\varphi^{(0,n)} = 0$ for all $n$. Therefore, the CS scalar field expansion takes the form:
\be
\label{th-ansatz}
\varphi= \epsilon \; \varphi^{(1,0)}(r,\theta) + \epsilon \;  \zeta \; \varphi^{(1,1)}(r,\theta) + \epsilon^{2} \; \varphi^{(2,0)}(r,\theta)\,.
\ee
Now we are ready to apply the procedure we prescribed above to solve the amended field equations, concentrating,  first, on the evolution equation of the CS scalar field.  Up to ${\cal{O}}(1,0)$, the evolution equation yields
\ba
\label{1st-eq}
k \varphi^{(1,0)}_{,rr} &+& \frac{2}{r} \varphi^{(1,0)}_{,r} \left( 1 - \frac{m}{r}+\frac{2r^2}{l^2} \right) + \frac{1}{r^{2}} \varphi^{(1,0)}_{,\theta\theta} + \frac{\cot{\theta}}{r^{2}} \varphi^{(1,0)}_{,\theta}
\nonumber \\
&=& - \frac{144 M^{3}}{r^{7}} \frac{\alpha}{\beta}  \frac{a}{m} \cos{\theta}\,.
\ea
In Eq. \eqref{1st-eq},  we have not taken into account  the potential $V(\vartheta)$. Solution of the  partial differential Eq. \eqref{1st-eq} consists of two parts, the  homogenous one, where the RHS of Eq \eqref{1st-eq} is zero, and the particular solution, where the RHS of Eq. \eqref{1st-eq} is not vanishing.  Therefore, the general solution is a superposition of the homogeneous  and a particular solution, i.e.,  $\varphi^{(1,0)} = \varphi^{(1,0)}_{H} + \varphi^{(1,0)}_{P}$. Now we are going to discuss the  homogeneous equation which is a separable one and yields:
\be
\varphi^{(1,0)}_{H}(r,\theta) = \Phi(r) \Phi(\theta).
\ee
The partial differential equation then becomes a set of ordinary differential equations for $\tilde{\varphi}$ and $\hat\varphi$,
that have the form:
\ba
\label{Hom-sol-1}
\Phi(r) &=& \tilde\varphi''(r)+\frac{2\tilde\varphi'(r)\left(\frac{r^2}{l^2}+1-\frac{m}{r}\right)}{rk}+\frac{c_1 \tilde\varphi(r)}{r^2l^2k}=0\,,\nonumber \\
\Phi(\theta) &=& \hat\varphi_{_{\theta \theta}}(\theta)+\hat\varphi_{_{\theta}}(\theta)\cot\theta-\frac{\hat\varphi(\theta)c_1}{l^2}=0\,,
\ea
where ${\displaystyle \hat\varphi_{_{\theta}}(\theta)=\frac{d\hat\varphi(\theta)}{d\theta}}$ and $c_{1}$ is an integration  constant  coming from the separation of variables. The solution of  $\hat\varphi(\theta)$ gives:
\ba
\label{Hom-sol-2}
\hat\varphi(\theta)=c_2LP\left(\frac{\sqrt{1-\frac{4c_1}{l^2}}-1}{2},\cos\theta\right)+c_3LQ\left(\frac{\sqrt{1-\frac{4c_1}{l^2}}-1}{2},\cos\theta\right)\,.
\ea
Here $LP(v, x)$ and $LQ(v, x)$ are  the Legendre and associated Legendre functions of first kind and both of them satisfy the differential equation:
\be
(1-x)^2y''(x)-2xy'(x)+v(v+1)y(x)=0\,.
\ee
The solution  $\hat\varphi(r) $ is not so easy to get  in an exact form, so we are going to derive an approximation form of the first  of Eqs. \eqref{Hom-sol-1} which yields:
\be \label{11}
\tilde\varphi''(r)+\tilde\varphi'(r)\left(\frac{1}{r}-\frac{1}{2m}-\frac{r}{4m^2}\right)-\left(\frac{c_1}{2ml^2r}+\frac{c_1}{4m^2l^2}\right)\tilde\varphi(r)=0\,.\ee
The solution of Eq. \eqref{11} takes the form:
\ba\label{solr}
&&\varphi(r)=c_4\,HB \left( 0,\sqrt {2},{\frac {2\,l^{2}-2 \,c_{{1}}}{l^{2}}},-{\frac { \left(l^{2} -2\,c_{{1 }} \right) \sqrt {2}}{l^{2}}},{\frac { \sqrt {2}r}{4m}} \right) +c_5\,HB \left( 0,\sqrt {2},{ \frac {2\,l^{2}-2\,c_{{1}}}{l^{2}}},-{\frac { \left(l^{2} -2\,c_{{1}} \right) \sqrt {2}}{l^{2}}},{\frac {\sqrt {2}r}{4m}} \right)\nonumber\\
&& \times\int \!\frac{e^{{\frac {r \left( 4\,m+r \right) }{8{m}^{2}}}}}{r \left[ HB \left( 0,\sqrt {2},{\frac {2\,l^{2}-2\,c_{{1}} }{l^{2}}},-{\frac { \left( -2\,c_{{1}}+l^{2} \right) \sqrt {2}}{l^{2}}},1/4\,{\frac {\sqrt {2}r}{m}} \right)  \right] ^2}{dr}
\,,
\ea
where $HB(\alpha,\beta,\gamma,\delta,z)$ is the Heun B function which is a solution of the differential equation \cite{Heun}:
\be
y''(z)-\frac{(z\beta-\alpha+2z^2-1)}z y'(z)-\frac{(2\alpha-2\gamma+4)z+\delta+\beta+\alpha \beta}{2z} y'(z)\,, \qquad y(0)=1, \qquad y'(z)=\frac{\delta+\beta+\alpha \beta}{2(1+\alpha)}\,.
\ee
Eq. \eqref{solr} gives no physical solution because it has no well defined behavior as $r\to \infty$, therefore we put $c_4=c_5=0$. Moreover, for finite energy $\hat\varphi(\theta)$  has a zero value which means $c_2=c_3=0$. The above discussion means that $\varphi^{(1,0)}_{H}(r,\theta) =$constant.
%
%

Since we derived  the homogeneous solution, we can proceed with  the particular solution. In this case, Eq. \eqref{1st-eq} can be rewritten as:
\be
 \tilde\varphi''(r)+\frac{2\tilde\varphi'(r)\left(\frac{r^2}{l^2}+1-\frac{m}{r}\right)}{rk}-\frac{2\tilde\varphi(r)}{r^2(1+\frac{r^2}{l^2}-\frac{2m}{r})}
 +\frac{144\alpha a m^2 }{\beta\,r^7(1+\frac{r^2}{l^2}-\frac{2m}{r})}=0\,.
\ee
In order to be capable of solving the above non-homogenous differential equation, we can write it in the asymptotic form as:
\be
 \tilde\varphi''(r)+2\tilde\varphi'(r)\left(\frac{2}{r}-\frac{l^2}{r^3}+\frac{3ml^2}{r^4}\right)-\frac{2l^2\tilde\varphi(r)} {r^4}
 +\frac{144\alpha a m^2 }{\beta\,r^7(1+\frac{r^2}{l^2}-\frac{2m}{r})}=0\,,
\ee
where we have put  $\varphi^{(1,0)}(r,\theta)=\cos\theta  \varphi^{(1,0)}(r)$.
The solution of the above differential equation takes the form:
\ba
\label{theta-sol-SR}
&&\varphi^{(1,0)}_{P} =
\frac{-144\,\cos\theta\,\alpha\,a\,
{l}^{2}{m}^{2}}{{\beta}{e^{ \left( 1-\frac{2\,m}{r}\right)\frac {
l^{2}}{{r}^{2}}}}}\biggl\{\int \! \frac{dr}{l^2r^3\left( \frac{{r}^{2}}{l^2}+1-\frac{2m}{r} \right)}
\int \!\frac{
e^{ \left( 1-\frac{2\,m}{r}\right)\frac {
l^{2}}{{r}^{2}}}}{r
^4}{dr}-\int \left[\!\int \!\frac{e^{ \left( 1-\frac{2\,m}{r}\right)\frac {
l^{2}}{{r}^{2}}}}{r^4}\right]{dr}\frac{1}{l^2r^3\left( \frac{{r}^{2}}{l^2}+1-\frac{2m}{r} \right)}{dr} \biggr\} \nonumber\\
&&\approx -\frac{6\alpha \,a}{5m^3\beta}\frac{r^2}{l^2}\left(1+\frac{15m}{8r}+\frac{44M^2}{15r^2}+\frac{7m^3}{r^3}\right)-\frac{\alpha \,a}{50m^3\beta}\left(1+\frac{150m}{r}
+\frac{439m^2}{r^2}+\frac{232m^3}{r^3}\right)\,.
\ea
In the above solution, we have  set the additional integration constant to zero since it does not give any contribute to the modified Einstein equations.

Since we derived the CS coupling scalar field, we can now search for the CS corrections to the metric perturbations.
It is worth  noticing that the stress-energy tensor of  the CS scalar, derived  here, enters the modified field equations at ${\cal{O}}(2,1)$, therefore,  it  has no impact on the metric perturbations. The modified Einstein equations can be separated into two categories:
the first one forms a closed system of partial differential equations
for $s^{(1,1)}$, $p^{(1,1)}$, $q^{(1,1)}$ and $n^{(1,1)}$, which constitute $(t,t)$, $(r,r)$, $(r,\theta)$, $(\theta,\theta)$ and $(\phi,\phi)$-components of the amended  Einstein equations. The second one consists of  a differential equation for $\omega^{(1,1)}$, more precisely the $(t,\phi)$-component of the amended Einstein equations.

The first category does not depend on the CS coupling field  $\varphi$, because it is exclusively  derived  from the Ricci tensor. We have that, using Eq.~\eqref{theta-sol-SR},
the output components of the C-tensor of the first category  is equal zero. Because the metric perturbations do not form a CS distortion (i.e.~they are  not depend on $\zeta$), therefore, we can put  them equal to zero, that is:  $s^{(1,1)}=0$, $p^{(1,1)}=0$, $q^{(1,1)}=0$ and $n^{(1,1)}=0$.

Therefore, the only non-vanishing  equation is the one  from the second group, which is:
\ba \label{14com}
&&12 k_1(r) r^2m\alpha\,\tilde\varphi_{r\theta} \left( r,\theta \right)+k_1(r)  {r}^{7}\sin \theta \omega_{rr} \left( r,\theta \right) -{r}^{5}\sin  \theta  \,\omega_{\theta \theta} \left( r,\theta \right)- 12k_1(r) r m\alpha\,\tilde\varphi_{\theta} \left( r,\theta \right) +{r}^{5} \bigg[ 4 k_1(r) r\sin \theta \omega_r \left( r,\theta \right)   \nonumber\\
 &&-3\omega_\theta \left( r,\theta \right)\cos\theta-6 \left\{ \frac{{r}^{2}}{l^{2}}\,\omega \left( r, \theta \right) + \left(\frac{2m}r -\frac{{r}^{2}}{l^2} \right) a \right\} \sin \theta \bigg]=0\, .
\ea
where $k_1(r)=\left[  \left(\frac{2m}{r}-1 \right)-\frac{{r}^{2}}{l^2} \right]$. Eq. \eqref{14com} coincides with what is derived in \cite{Yunes:2009hc} when $\Lambda=0$ or $l=\infty$.  Using Eq. \eqref{theta-sol-SR} in \eqref{14com}, we get the most general solution which is a linear combination of a homogeneous solution and a particular
one. With the same discussion of the CS coupling scalar field on the homogenous  equation, we can show that the homogenous solution of Eq. \eqref{14com} is not a physical one and therefore, we will not discuss it. The particular solution  of Eq. \eqref{14com} is given by:
\be
\label{w-sol-SR}
\omega^{(1,1)} = - \frac{5}{8} \frac{\alpha^{2}}{\beta \kappa} \frac{a}{r^{6}}\left( 1 + \frac{12}{7} \frac{m}{r} + \frac{27}{10} \frac{m^{2}}{r^{2}} \right)+\frac{528}{125} \frac{\alpha^{2}}{\beta l^2 \kappa} \frac{a}{r^{4}}\left( 1 + \frac{525}{176} \frac{m}{r}  \right)\,.
\ee
 We stress here  that this perturbation is  proportional to $\zeta$ and it possesses the correct units $[\omega] = L^{-1}$, because $[\xi] = L^{4}$.

Thus the full gravitomagnetic metric perturbation up to the linear order in $\zeta$ and $\epsilon$ is
\be\label{OMG}
\omega \approx-\frac{a}{l^{2}}+\frac{2m a}{r^3}+\frac{528}{125} \frac{\alpha^{2}}{\beta l^2 \kappa} \frac{a}{r^{4}}\left( 1 + \frac{525}{176} \frac{m}{r}  \right)-\frac{5}{8} \frac{\alpha^{2}}{\beta \kappa} \frac{a}{r^{6}}\left( 1 + \frac{12}{7} \frac{m}{r} + \frac{27}{10} \frac{m^{2}}{r^{2}} \right)\,.
\ee
Eq.  \eqref{w-sol-SR} is useful to construct the first slow-rotating (A)dS BH solution in the dynamical CS amended gravity. We stress  that the perturbation is slightly suppressed in the asymptotic  field which is decaying as $r^{-4}$, that supposes that its effect can be felt  in the strong field region. The above result shows that the contribution of the (A)dS spacetime makes the BH  stronger when  compared with a BH without (A)dS spacetime.

As expected, the correction of the metric is a small $\xi$-distortion of the (A)dS-Kerr metric. This result is consistent  with the small-coupling approximation.
 We  can show that this approximation is consistent by evaluating the next leading order correction to $\varphi$. This correction   consists of  $\varphi^{(2,0)}$ and $\varphi^{(1,1)}$, which can be calculated by solving the evolution equation to the next order. We get:
\be \label{theta_11}
\varphi^{(2,0)}=0, \qquad \qquad
\varphi^{(1,1)} \approx- \frac{3}{2} \frac{\alpha}{\beta} \frac{\xi a}{m^2} \frac{\cos{\theta}}{ r} \left(1 + \frac{4 m}{3r} + \frac{2 m^{2}}{r^{2}} + \frac{42m^{3}}{15 r^{3}}\right)-\frac{3}{2} \frac{\alpha}{\beta} \frac{\xi a}{l^2} \frac{\cos{\theta}}{ r}\,.
\ee
If we  use this improved
$\varphi$ solution in the CS modified field equations, we get a correction  proportional to $\zeta^{2} \epsilon$, which we are neglecting.

\section{Properties of the   (A)dS rotating Solution}
\label{properties}

We  are going to study now some  geometric properties of the slowly (A)dS-rotating solution.
The non-vanishing
metric components are:
\ba\label{sol:metric_elements}
g_{tt} &=& -k  -\frac{\left(\frac{r^{2}}{l^2}\sin^2\theta+\frac{2m}{r}\cos^2\theta\right)a^2}{r^{2}} \nonumber, \\
g_{t\phi} &=& \left(\frac{r^{2}}{l^2}-\frac{2m}{r}\right) a\,\sin^2\theta
+\frac{5}{8} \frac{\alpha^{2}}{\beta \kappa} \frac{a}{r^{4}}\left( 1 + \frac{12}{7} \frac{m}{r} + \frac{27}{10} \frac{m^{2}}{r^{2}} \right)\,\sin^2\theta-\frac{528}{125} \frac{\alpha^{2}}{\beta l^2 \kappa} \frac{a}{r^{2}}\left( 1 + \frac{525}{176} \frac{m}{r}  \right)\,\sin^2\theta\,,
\nonumber \\
g_{rr} &=& \frac{1}{k} +  \frac{a^{2}\left[r(1+\frac{r^2}{l^2})\sin^2\theta+\frac{2m}{r}\cos^2\theta\right]}{\frac{r^{2}}{l^2}k}\,,
\nonumber \\
g_{\theta \theta} &=& r^{2} + \frac{a^{2}}{r^{2}} \left(1+\frac{r^2}{l^2}\right)\cos^{2}{\theta}\,,
\nonumber \\
g_{\phi \phi} &=& r^{2} \sin^{2}{\theta}-a^2\sin^2\theta \left(1+\frac{r^2}{l^2} + \frac{2 m}{r} \sin^{2}{\theta} \right)\,,
\ea
which are correct up to orders ${\cal{O}}(2,0)$, ${\cal{O}}(1,1)$ and ${\cal{O}}(0,2)$.
A question could be if the CS corrections can be gauged away using a coordinate
transformation. This situation  is not possible for the following reasons.
The curvature invariants  indicate  that  the CS corrected solution is of
 the order reported here. From the most evident  of these invariants, the Pontryagin density
$\pont$, we see that it is proportional to $\square \vartheta$, and thus the shift from the (A)dS-Kerr solution
can be easily calculated from (\ref{theta_11}). Furthermore, as we will see soon,
the location of the inner-most stable circular orbit is CS corrected. This fact  points out
 that the CS amended is a non-trivial geometric perturbation of (A)dS-Kerr. { In Appendix~\ref{Sec:thermodynamics} we give another physical application which is related  to the thermodynamical properties of the BH of this study.}

\section{Geodesics of the (A)dS-slowly rotating black hole}\label{geo}

In this section, we are going  to consider the geodesics of a test particle in the (A)dS-slowly  rotating BH background assuming the orbits of the particle in the equatorial plane with $\theta=\frac{\pi}{2}$. In  the case of equatorial plane, the  metric reduces  to the following form:
\begin{eqnarray}
 ds^{2}=-kdt^{2}+k_1dr^{2}+r^2d\phi^{2}-2 \omega\,r^2\,\sin^2\theta dtd\phi,
\end{eqnarray}
where ${\displaystyle k = \frac{r^2}{l^2}+1 - \frac{2 M}r}$, ${\displaystyle k_1=\frac{1}{k}}$,  and $\omega$ is given by Eq. \eqref{OMG}.

This (A)dS-slowly rotating BH background has two Killing fields $\partial_{t}$ and $\partial_{\phi}$. Therefore, there are two constants  $L$ and $E$, that are the orbital angular momentum and energy conserved quantities per unit mass. Using the momentum $p_{\mu}=g_{\mu\nu}\dot{x}^{\nu}$, we get:
\begin{eqnarray}\label{Lan}
  L&=&p_{\varphi}=-\omega\dot{t}+r^2\dot{\phi}\,,\\\label{En}
  -E&=&p_{t}=-k\dot{t}-\omega\dot{\phi}\,.
\end{eqnarray}
From Eqs. \eqref{Lan} and \eqref{En}, we obtain the   $\phi$-motion and $t$-motion as:
\begin{eqnarray}
  \dot{\varphi}=\frac{E\omega+L k}{\omega^{2}+r^2k}\,, \qquad \qquad \dot{t}=\frac{Er^2-L \omega}{\omega^{2}+r^2k}\,.
\end{eqnarray}
The normalizing condition,  $g_{\mu\nu}\dot{x}^{\mu}\dot{x}^{\nu}=-\delta^{2}$, gives the $r$-motion as:
\begin{eqnarray}
 \dot{r}^{2}=\frac{r^2E^{2}-L(2\omega\,E+kL)}{k_1(\omega^{2}+r^2k)}-\frac{\delta^{2}}{k_1}\,,
\end{eqnarray}
where $\delta^{2}=1$ and $0$ for  time-like  and null geodesics, respectively. Thus,  the $r$-motion can be rewritten as:
\begin{eqnarray}
 \dot{r}^{2}+V_{\textrm eff}=0\,,
\end{eqnarray}
where the effective potential $V_{\textrm eff}$ is given by: \be V_{\textrm eff}=\frac{L(2\omega\,E+kL)-r^2\,E^{2}}{k_1(\omega^{2}+r^2 k)}+\frac{\delta^{2}}{k_1}\,.\ee
\begin{figure*}
\centering
\subfigure[~Effective potential of GR]{\label{fig:3a}\includegraphics[scale=0.2]{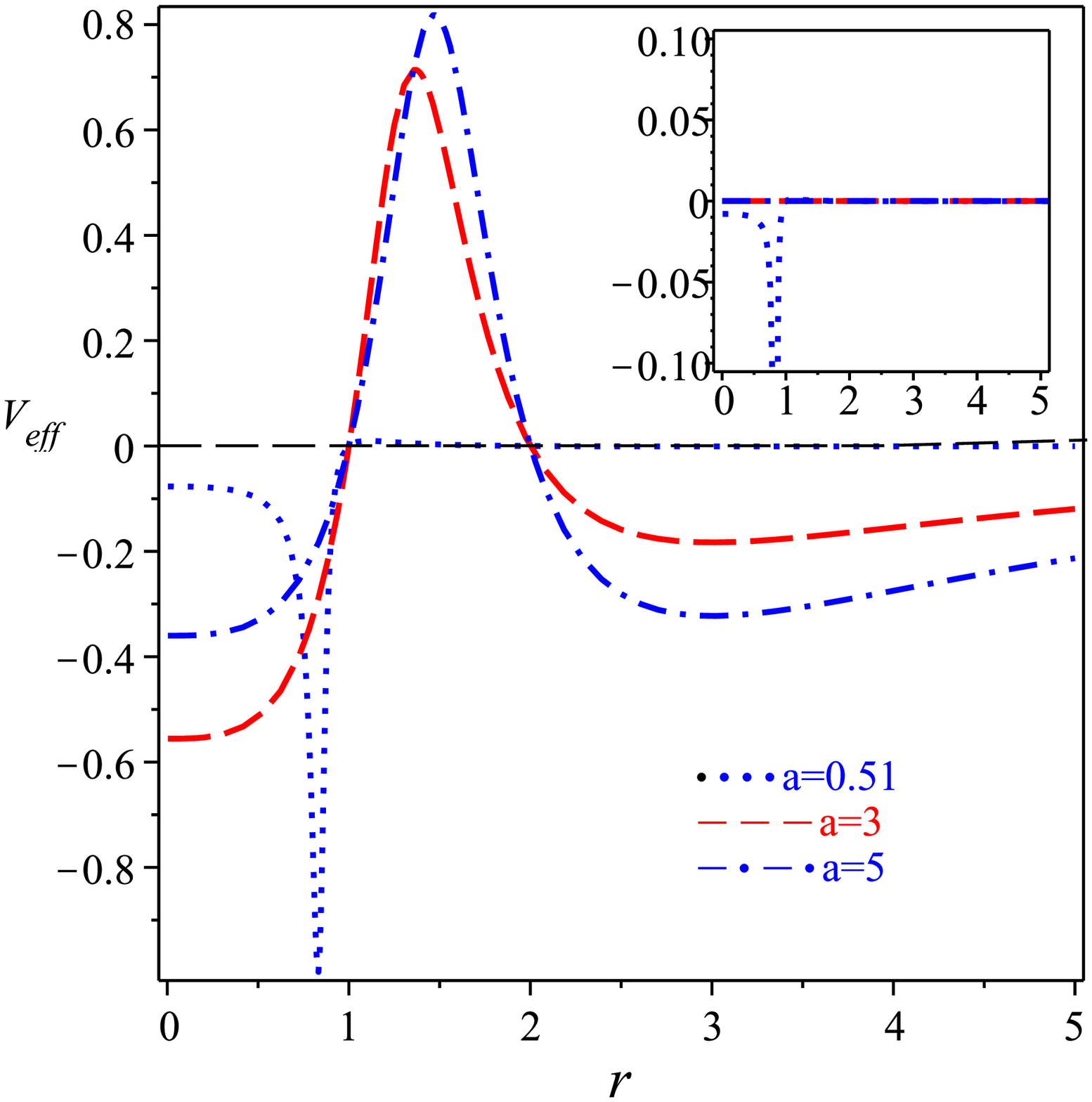}}\hspace{1cm}
\subfigure[~Effective potential of CS when $\alpha$ has a constant value]{\label{fig:3b}\includegraphics[scale=0.2]{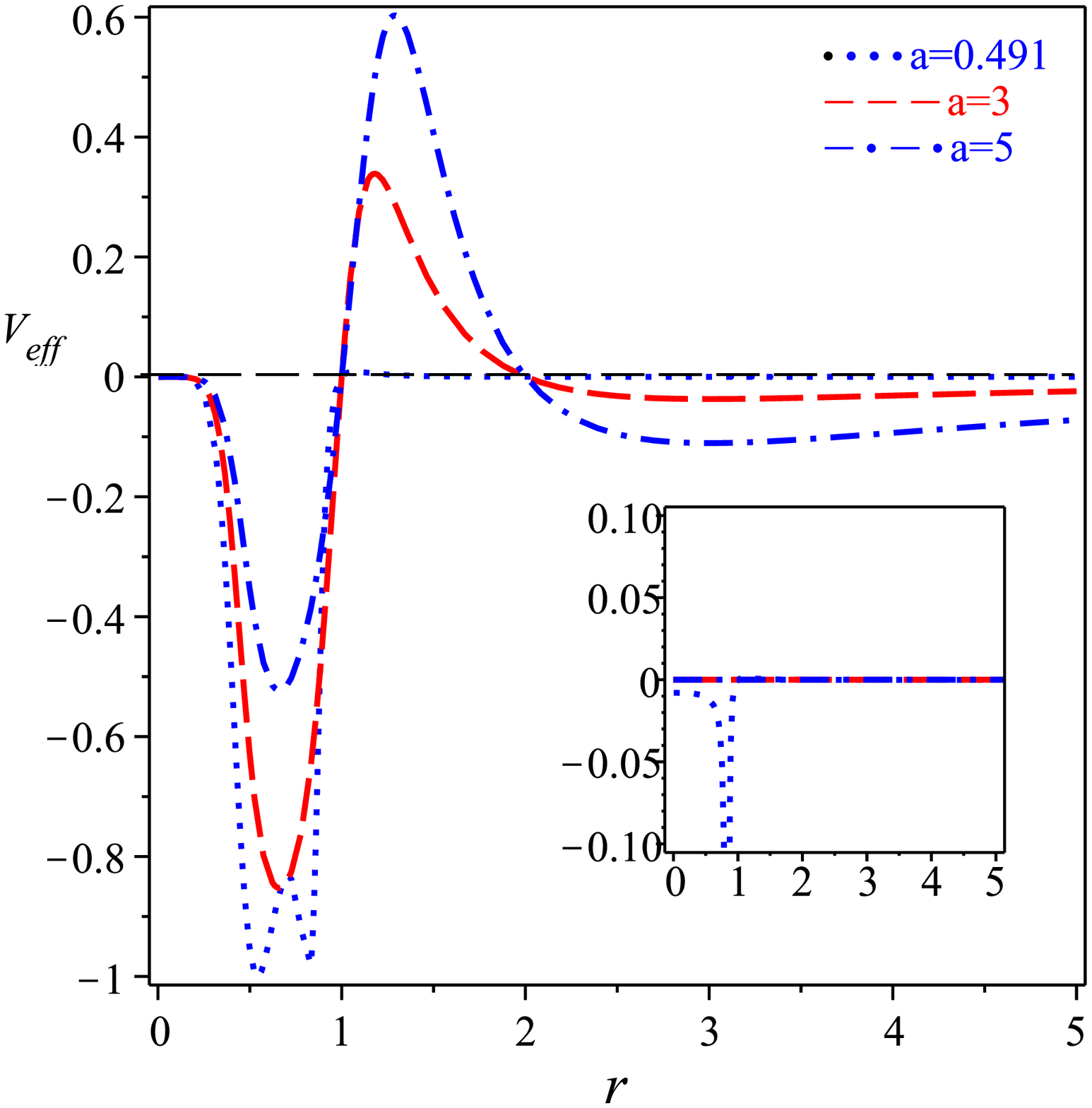}}\hspace{1cm}
\subfigure[~Effective potential of CS when the rotation parameter, $a$, has a constant value]{\label{fig:3c}\includegraphics[scale=0.2]{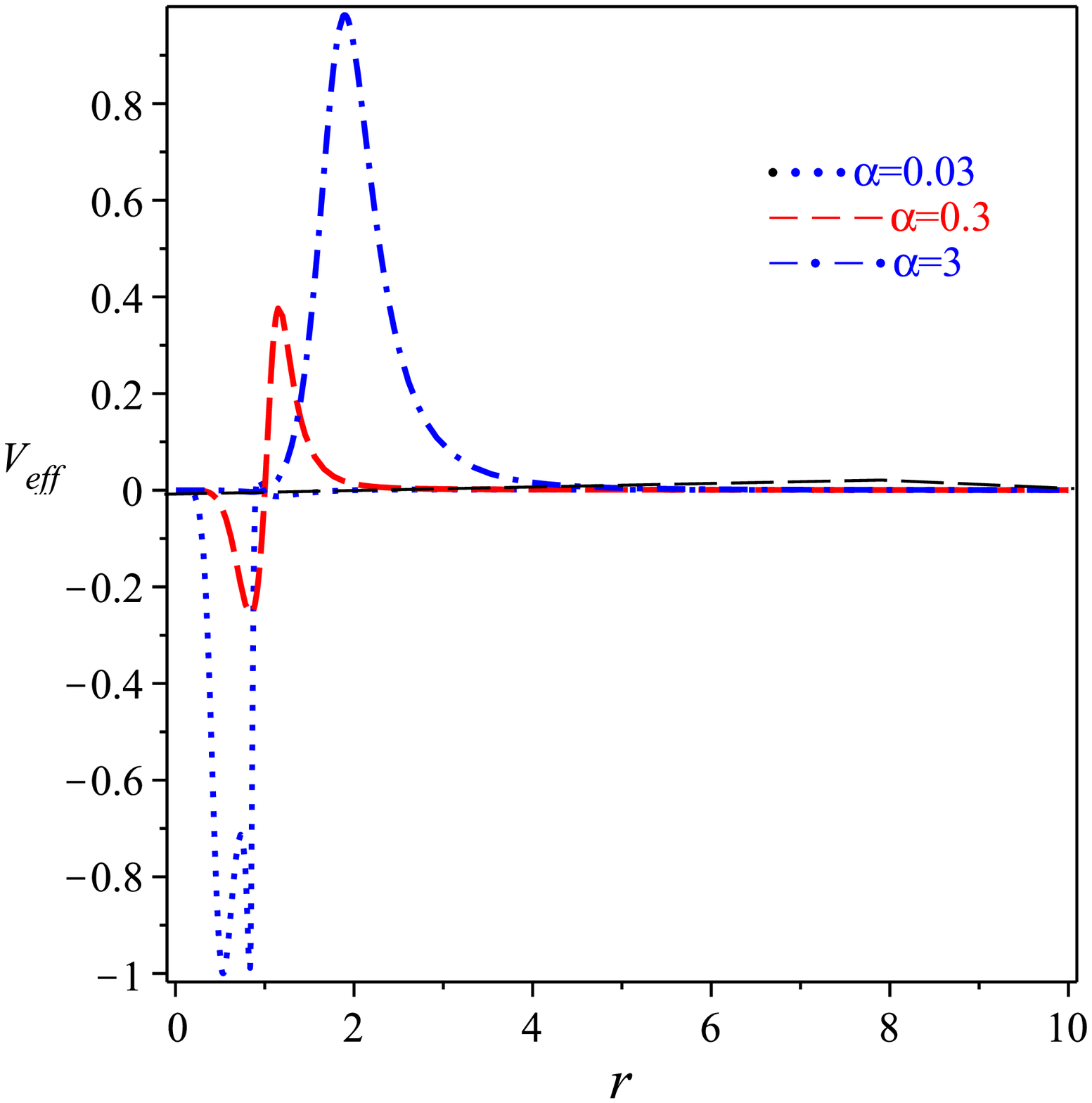}}\hspace{1cm}
\caption{Schematic plots of the effective potential $V_{eff}$. \subref{fig:3a} represents the effective potential of GR using different numerical values of the rotation parameter $a$, while \subref{fig:3b} represents the effective potential of CS  using different numerical values of the rotation parameter, and \subref{fig:3c} represents the effective potential of CS  using different numerical values of the  dimensional parameter $\alpha$. All the above figures are plotted for $M=1$, $l=1$, $L=1$, $\xi=1$, $\kappa=1$, and $\beta=1$.}
\label{Fig:3}
\end{figure*}
In Fig. \ref{Fig:3}, we plot  the behavior of the effective potential for a photon using different values of the parameters. Because kinetic energy is always positive, i.e.,  $\dot{r}^{2}>0$, therefore, the ranges of  negative effective potential are allowed for the photon  traveling. For fixed $M=1$, $l=1$, $L=1$, $\xi=1$, $\kappa=1$, and $\beta=1$, we can find that there exists a peak. In the GR case, with the increase of  rotation $a$, the peak increases and approaches to zero at  certain radius $r$. Then further increasing $a$, the peak is positive, and thus it  prevents the photons from outside to fall into the BH.  The same analysis can be applied for the CS  case with the increase of the rotation $a$: The peak increases and approaches to zero at certain radius $r$. Then further increasing $a$, the peak is positive, and thus it  prevents the photons from outside to fall into the BH. However for the CS case and for increasing the dimensional parameter $\alpha$, the peak is positive, and thus it prevents the photons from outside to fall into the BH.

Interesting phenomena occur when the peak has a zero value. In this situation, the photon acquires a vanishing radial velocity  with a non-vanishing transverse velocity. Thus, the photon  circles the BH one loop by one loop. Such an unstable circular photon orbit is determined by
\begin{eqnarray}
 V_{\textrm eff}(r_{\textrm o}, L_{\textrm o})=0, \quad
 \partial_{r}V_{\textrm eff}(r_{\textrm co}, L_{\textrm o})=0.
\end{eqnarray}
Solving these conditions, we can obtain the radius $r_{\textrm co}$ and the angular momentum $L_{\textrm co}$ for the circular orbit. Plugging the effective potential into the above two conditions, we have
\begin{eqnarray}
 &&L_{\textrm o}=\frac{-\omega(r_{\textrm o})+\sqrt{r_{\textrm o}^2k(r_{\textrm o})+\omega(r_{\textrm o})^{2}}}{k(r_{\textrm o})},\label{c1}\\
 &&2L_{\textrm o}\big[k(r_{\textrm o})\omega'(r_{\textrm o})-k'(r_{\textrm o})\omega(r_{\textrm o})\big]-r_{\textrm o}\big[2k(r_{\textrm o})-r_{\textrm o}k'(r_{\textrm o})\big]=0.\label{c2}
\end{eqnarray}
The prime indicates the derivative with respect to $r$. Clearly, this analysis can be extended in detail to obtain the so called BH shadow \cite{Wei:2019pjf, Falcke}.

\section{Discussion and Conclusions}
\label{conclusions}

In this work,  we discussed the  (A)dS-slowly rotating BH
in the dynamical CS modified gravity, up to the leading order in the coupling constant.
Due to the fact that the non-dynamical case of CS modified gravity does not provide any new solution different from GR, we considered only the dynamical case of CS modified gravity. Specifically,  we  considered  physically reasonable
conditions that make the CS scalar field $\varphi$  obey the symmetries
of the spacetime, and has finite, positive energy exterior to the BH
event horizon. The resulting  BH solution describes an inherently {\it strong-field}
perturbation of (A)dS-Kerr, where  deformation of the background geometry
decays as $(1/r)^2$, compared with the slowly rotating BH solution where it decays as $(1/r)^4$. As a result, the deformation is coherent with  weak-field
bounds, while  very different
phenomena can emerge in strong-field scenarios involving spinning BHs. In particular, they could be investigated considering compact object mergers, inner edges of accretion disks, gravitational
collapse, and so on.

Concerning the  structure  of the BH solution, we revealed  that, up to the leading order of
perturbation, the  ergosphere locations and horizon are unchanged if compared to the (A)dS-Kerr BH solution.
Additionally also the ADM mass and the angular momentum
of the (A)dS-Kerr spacetime result unaltered.  In particular, the CS scalar field $\varphi$ consists of two terms: The first  depends on the (A)dS Planck-scale, the second  depends on the rotation parameter $a$. The scalar field cannot reduce to the one derived in \cite{Yunes:2009hc} due to the contribution of the  Planck-scale.  Up to the leading order,  we have show that all the thermodynamical quantities do not  feel  the  CS effect.

Furthermore, for fixed  values of the parameters  characterizing the model, we have shown that there exists a peak in the effective potential $V_{eff}$. For the case of GR,  it is possible to  show that,  with increasing  value  of the rotation $a$, the peak increases and approaches to zero at  certain radius $r$, then, further increasing  the rotation,  the peak becomes positive, and thus   prevent the photons from outside to fall into the BH.  On the other hand,  for the CS  case, we show that increasing the value of the rotation, the peak increases and it approaches to zero at  certain radius $r$. Then,  increasing more  the rotation parameter,   the  peak  becomes positive, and thus it  prevents the photons from outside to fall into the BH. Same behavior emerges for  the CS case. In this case,  increasing the dimensional parameter $\alpha$, the peak becomes positive, and thus, again,  it  prevents  photons from outside to fall into the BH.

The study of CS case has a special  importance for the fast growing
field of gravitational wave of astronomy. On one hand gravitational wave  detectors, like space-based detector LISA (Laser Interferometer
Space Antenna)~\cite{Danzmann:2003tv} or  LIGO (Laser Interferometer Gravitational Observatory)~\cite{Abramovici:1992ah,Waldmann:2006bm,LIGOScientific:2007fwp}
 could be
useful  to put  constraints on the CS modifications with respect to GR.  As expected, such modifications could emerge in strong field regime so that gravitational waves could be a formidable tool to discriminate among possible BH solutions and then concurring theories of gravity.

On the other hand, a detailed geodesic analysis can give contributions in discriminating BH solution by a careful reconstruction of the BH shadow. In a forthcoming paper, we will consider this approach taking into account available observational data.
\bigskip
\appendix
\section{Thermodynamics of (A)dS-slow rotating black hole}\label{Sec:thermodynamics}
It is well known that thermodynamics can be extremely useful to fix global features of BH. In particular when exotic fluids or perturbations are considered. See for example \cite{Capozziello:2022ygp}.

\begin{figure}
  \centering
  \includegraphics[scale=0.3]{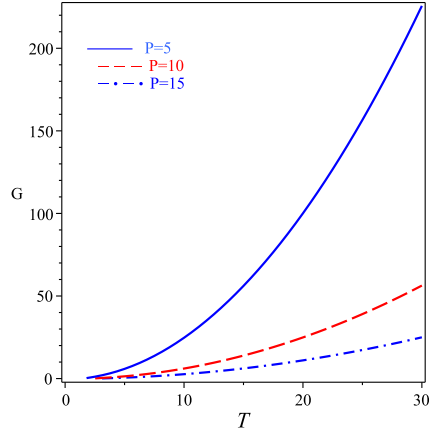}
  \caption{The Gibbs free energy for the Kerr-AdS BH  for various values of the pressure $P$.The horizon radius increases from left to right along the curve, and a sudden change of the horizon radius occurs in the small-large BH phase transition point.}
  \label{Fig: Gibbs}
\end{figure}

Here, the event horizon radius $r_{\text{h}}$ of the (A)dS-slowly rotating BH is fixed by the equation
\begin{equation}\label{horizonD}
  1+\frac{r^2}{l^2}-\frac{2m}{r}=0 \Rightarrow \rh={\frac { \left( 27\,m{l}^{2}+3\,\sqrt {3}\sqrt {{l}^{4} \left( 27\,{m}
^{2}+{l}^{2} \right) } \right) ^{2/3}-3\,{l}^{2}}{3\sqrt [3]{27\,m{l}^{
2}+3\,\sqrt {3}\sqrt {{l}^{4} \left( 27\,{m}^{2}+{l}^{2} \right) }}}}\,.
\end{equation}
The angular velocity
 of the observer, moving on the orbits of constant
$r$ and $\theta$, turns out to be
\begin{equation}\label{angul1}
  \Omega=-\frac{g_{t\,\phi}}{g_{\phi\, \phi}}=-\frac{a}{l^{2}}+\frac{2m a}{r^3}+\frac{528}{125} \frac{\alpha^{2}}{\beta l^2 \kappa} \frac{a}{r^{4}}\left( 1 + \frac{525}{176} \frac{m}{r}  \right)-\frac{5}{8} \frac{\alpha^{2}}{\beta \kappa} \frac{a}{r^{6}}\left( 1 + \frac{12}{7} \frac{m}{r} + \frac{27}{10} \frac{m^{2}}{r^{2}} \right)\,.
\end{equation}
In the case of rotating regular (A)dS BH,  the angular velocity \eqref{angul1} does not vanish at
asymptotic infinity and it gives the finite quantity:
\begin{equation}\label{angul2}
  \Omega_{\infty}=-\frac{a}{l^{2}}.
\end{equation}
The (A)dS-slowly rotating BH metric describes a rotating spacetime with the following angular velocity at the event horizon
\begin{equation}\label{angul}
  \Omega_{\text{h}}=-\frac{a}{l^{2}}+\frac{2m a}{r_h{}^3}+\frac{528}{125} \frac{\alpha^{2}}{\beta l^2 \kappa} \frac{a}{r_h{}^{4}}\left( 1 + \frac{525}{176} \frac{m}{r_h{}}  \right)-\frac{5}{8} \frac{\alpha^{2}}{\beta \kappa} \frac{a}{r_h{}^{6}}\left( 1 + \frac{12}{7} \frac{m}{r_h{}} + \frac{27}{10} \frac{m^{2}}{r_h{}^{2}} \right)\,.
\end{equation}
Throughout the extended phase space, the mass of the BH is explained as enthalpy instead of internal energy~\cite{Kastor:2009wy,Nashed:2016tbj}. The BH mass and angular momentum  have been first calculated using the Hamiltonian procedure  derived from the generators of $SO(3,2)$~\cite{Gibbons:2004ai,Henneaux:1985tv,Nashed:2018cth}:
\begin{equation}\label{Mass1}
   M=m,  \qquad J=ma.
\end{equation}
The above expressions are linked to the mass parameter $m$ and angular parameter $a$.
The Hawking temperature of the (A)dS-slowly rotating BH is given by:
\begin{equation}\label{HawkTemp}
  T=\frac{3\rh^2+l^2}{4\pi\,\rh l^2}\,,
\end{equation}
and the entropy takes the form
\begin{equation}
  S=\pi\rh^2.
\end{equation}
In the extended phase space, the negative cosmological constant can be interpreted as a pressure~\cite{Kastor:2009wy}
\begin{equation}
  P=\frac{3}{8\pi}\frac{1}{l^2},
\end{equation}
and  the conjugate thermodynamic variable is:
\begin{equation}
  V=\frac{4 \pi \rh^3}{3},
\end{equation}
 with the specific volume
\begin{equation}\label{SpV}
  v=2\rh.
\end{equation}
Using the above thermodynamic variables, we can easily satisfy the first law of (A)dS slowly rotating BH thermodynamics and the Smarr relation as:
\ba
  dM = TdS+VdP, \label{1srlaw} \\
  M = 2TS-2PV.
\ea
The Gibbs free energy of (A)dS-slowly rotating BH is given by
\ba
 G =M-T S \\
              =\frac{\rh(l^2-\rh^2)}{4l^2} \,.\label{Gibbs}
\ea
The Gibbs free energy $G$ and temperature $T$ can be expressed in terms of pressure $P$, entropy $S$, and angular momentum $J$ as \cite{Wei:2015ana}:
\ba
  G = \frac{\sqrt{ S}(3-8PS)}{12\sqrt{\pi}} ,\\
  T= \frac{(8PS+1)}{4\sqrt{\pi\,S}}.
\ea
We plot the Gibbs free energy as a function of the BH temperature $T$ with fixed angular momentum $J$ and various thermodynamic pressure $P$  in Fig.~\ref{Fig: Gibbs}.
As Fig.~\ref{Fig: Gibbs} shows, when the  pressure decreases,  the value of the Gibbs free energy increases.

\begin{figure}
  \centering
  \includegraphics[scale=0.3]{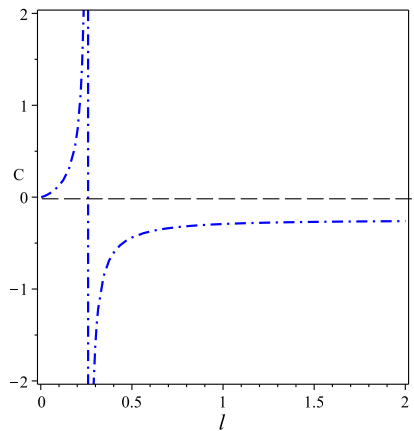}
  \caption{The heat capacity of (A)dS-slowly rotating BH.}
  \label{Fig:Heat}
\end{figure}

Another test to check the thermodynamic stability of a BH is given
by its heat capacity content. When the BH has a  positive heat capacity,
it is locally thermodynamically stable; whereas a
negative heat capacity shows thermodynamic instability. The heat capacity is defined as:
\begin{equation}\label{heatcapacity}
{\mathcal C} =\frac{\partial M}{\partial r_h}\left(\frac{\partial T}{\partial r_h}\right)^{-1}\,.
\end{equation}
Using Eqs. (\ref{Mass1}) and  (\ref{HawkTemp}) into Eq. (\ref{heatcapacity}) we get:
\ba\label{heatc}
&&{\mathcal C} =\frac{2}{9}\, \bigg( 9\,\sqrt [3]{27\,M{l}^{2}+3\,\sqrt {3}\sqrt {{l}^
{4} \left( 27\,{M}^{2}+{l}^{2} \right) }}M{l}^{2}+\sqrt [3]{27\,M{l}^{
2}+3\,\sqrt {3}\sqrt {{l}^{4} \left( 27\,{M}^{2}+{l}^{2} \right) }}
\sqrt {3}\sqrt {{l}^{4} \left( 27\,{M}^{2}+{l}^{2} \right) }\nonumber\\
&&-{l}^{2}
 \left( 27\,M{l}^{2}+3\,\sqrt {3}\sqrt {{l}^{4} \left( 27\,{M}^{2}+{l}
^{2} \right) } \right) ^{2/3}+3\,{l}^{4} \bigg) \pi\, \left(  \left(
27\,M{l}^{2}+3\,\sqrt {3}\sqrt {{l}^{4} \left( 27\,{M}^{2}+{l}^{2}
 \right) } \right) ^{2/3}-3\,{l}^{2} \right) ^{2}\nonumber\\
&&\bigg\{ \left( 27\,M{l}^{2
}+3\,\sqrt {3}\sqrt {{l}^{4} \left( 27\,{M}^{2}+{l}^{2} \right) }
 \right) ^{2/3} \bigg[ 9\,\sqrt [3]{27\,M{l}^{2}+3\,\sqrt {3}\sqrt {{l
}^{4} \left( 27\,{M}^{2}+{l}^{2} \right) }}M{l}^{2}+\sqrt [3]\nonumber\\
&&{27\,M{l}
^{2}+3\,\sqrt {3}\sqrt {{l}^{4} \left( 27\,{M}^{2}+{l}^{2} \right) }}
\sqrt {3}\sqrt {{l}^{4} \left( 27\,{M}^{2}+{l}^{2} \right) }-3\,{l}^{2
} \left( 27\,M{l}^{2}+3\,\sqrt {3}\sqrt {{l}^{4} \left( 27\,{M}^{2}+{l
}^{2} \right) } \right) ^{2/3}+3\,{l}^{4} \bigg] \bigg\}^{-1}\,.
\ea
Here we define the degenerate horizon as
\be\label{deg}
r_d=\frac{\partial M}{\partial r_h}\,.
\ee
Using Eq. (\ref{Mass1}) in Eq.  (\ref{deg}) we get
\ba
&&\frac{3r^2_{r_h}+l^2}{2l^2}\nonumber\\
&&\equiv \frac{1}{2}\bigg(9\,\sqrt [3]{27\,M{l}^{2}+3\,\sqrt {3}\sqrt {{l}^{4}
 \left( {l}^{2}+27\,{M}^{2} \right) }}M{l}^{2}+\sqrt [3]{27\,M{l}^{2}+
3\,\sqrt {3}\sqrt {{l}^{4} \left( {l}^{2}+27\,{M}^{2} \right) }}\sqrt
{3}\sqrt {{l}^{4} \left( {l}^{2}+27\,{M}^{2} \right) }\nonumber\\
&&- \left( 27\,M{l
}^{2}+3\,\sqrt {3}\sqrt {{l}^{4} \left( {l}^{2}+27\,{M}^{2} \right) }
 \right) ^{2/3}{l}^{2}+3\,{l}^{4}\bigg)\bigg\{ \left( 27\,M{l}^{2}+3\,\sqrt {3}
\sqrt {{l}^{4} \left( {l}^{2}+27\,{M}^{2} \right) } \right) ^{2/3}{l}^
{2}\bigg\}^{-1}
\,.
\ea
The behavior of Eq.\eqref{heatc} is shown in Fig. ~\ref{Fig:Heat}.
As Fig. ~\ref{Fig:Heat} shows,  the heat capacity depends on $r_d$ in which, for $l<r_d$, we have a negative heat capacity and vice-versa.

 \section*{Acknowledgements}
 This paper is based upon work from COST Action CA21136 {\it Addressing
observational tensions in cosmology with systematics and fundamental physics} (CosmoVerse) supported by COST (European Cooperation in Science and Technology).
 SC acknowledges the Istituto Nazionale di Fisica Nucleare (INFN) Sez. di Napoli,  Iniziative Specifiche QGSKY and MOONLIGHT,   and the Istituto Nazionale di Alta Matematica (INdAM), gruppo GNFM.



\end{document}